%% file: main.tex
\documentclass{article}

\usepackage{amsmath}
\usepackage{amssymb}
\usepackage{graphicx}
\usepackage[ruled,vlined]{algorithm2e}

\usepackage{glossaries}
\usepackage{subcaption}

\newglossaryentry{quars}{
    name={QuaRs},
    description={Quantile Reshuffling}
}

\begin{document}


\title{QuaRs: A Transform for Better Lossless Compression of Integers}
\author{Jonas G. Matt}
\date{\today}
\maketitle

\section{Introduction}\label{sec:introduction}

The rise of integer-valued data across various domains, particularly with the proliferation of the Internet of Things (IoT), has spurred significant interest in efficient integer compression techniques \cite{mogahedDevelopmentLosslessData2018,spiegelComparativeExperimentalStudy2018,vestergaardTitchyOnlineTimeSeries2021,deoliveiraTimeSeriesCompression2023,hughesComparisonLossyLossless2023}.
These techniques aim to reduce storage and transmission costs, which are critical for managing the vast amounts of data generated in modern applications.
While lossy compression methods, such as downsampling and quantization, are effective in reducing data size by sacrificing precision, lossless compression remains indispensable for applications that demand exact data recovery.
Moreover, lossless compression can often complement lossy methods, further enhancing overall efficiency.\par
Existing integer compression methods typically prioritize speed over compression ratio and are commonly built on the ``smaller-numbers-less-bits'' principle \cite {golombRunlengthEncodingsCorresp1966,eliasUniversalCodewordSets1975,apostolicoRobustTransmissionUnbounded1987,lemireDecodingBillionsIntegers2015,al-kadhimEnergyEfficientData2021,blalockSprintzTimeSeries2018}.
This principle is based on the assumption that smaller numbers (in absolute value) occur more frequently and can therefore be encoded using fewer bits.
Implicitly, this assumes that the numerical distribution of the data is unimodal around $0$.\par
However, in practical scenarios, this assumption often fails.
For example, numeric data may exhibit multimodal distributions, characterized by multiple distinct peaks, or sparse distributions, characterized by a spread-out set of frequent values.
Such cases challenge the efficacy of compression methods reliant on the smaller-numbers-less-bits principle, leading to suboptimal performance.\par
In this manuscript, we introduce \gls{quars}, a novel transformation designed to enhance the compression achieved by fast integer compression techniques.
\gls{quars} reshapes arbitrary numerical distributions into unimodal distributions centered around zero, making them amenable to efficient compression.
This is achieved by remapping data bins based on quantiles, such that more frequent values are assigned smaller absolute magnitudes, while less frequent values are mapped to larger ones.\par
The transformation is fast-to-compute, invertible, and seamlessly integrates with existing compression methods.
\gls{quars} enables effective and fast compression of data distributions that deviate from the assumptions underlying conventional methods.
The computational complexity of QuaRs is low: practically, it requires only $\mathcal{O}(N \log N)$ operations to compute the quantiles and $\mathcal{O}(N)$ operations to apply the transformation to a numeric data set (e.g. a time series) of size $N$, i.e., containing $N$ integers.

\section{QuaRs: Quantile Reshuffling}\label{sec:quars}

This section describes \gls{quars}, the proposed transformation designed to enhance the compression of integer data.
\gls{quars} transforms any given numerical distribution into a unimodal distribution centered around zero, making the underlying data more amenable to compression with certain methods.
It accomplishes this by remapping data bins such that more frequent values are assigned smaller absolute magnitudes, while less frequent values are mapped to larger ones.
The data's sample quantiles are used to define the bins, making the choice of bins dependent on the input data.\par
The only tunable parameter of \gls{quars} is the total number of quantiles, $q$, which controls the granularity of the transformation.
The number of bins effectively used for the transformation can be smaller (if some quantile values occur more than twice) but is never larger than $q+1$.
Detailed procedures for encoding and decoding with \gls{quars} are provided in Algorithms \ref{alg:quars-encode} and \ref{alg:quars-decode}, respectively.

\input{quars_algorithm.tex}

The encoding stage takes as input a data set $\mathcal{D}$, consisting of $N$ integers, and an integer $q > 0$, representing the number of quantiles to partition the data into.
It outputs the sorted representation of the bins used for transformation, $\mathcal{B}^*$, and the QuaRs-transformed data $\mathcal{D}_\text{QuaRs}$.
The algorithm first divides the dataset $\mathcal{D}$ into $q$ quantiles.
Quantiles are calculated by sorting the integers in $\mathcal{D}$ in ascending order and identifying $q-1$ evenly spaced threshold values.
These thresholds partition the data into approximately equal-sized subsets.
For example, if $q=4$ the thresholds might correspond to the 25th, 50th, and 75th percentiles.\\
Based on the computed quantiles, the bins ($\mathcal{B}$) are constructed.
Each bin represents a range of values within $\mathcal{D}$.
Additional adjustments are made to ensure all unique thresholds are included: Duplicate quantiles are replaced with slightly larger values and the maximum value of $\mathcal{D}$ is always included as the upper bound of the last bin.
As a result, the bins cover the entire range of $\mathcal{D}$.\par
The bins are then sorted based on two criteria: the width of the bin (increasing) and the number of items in the bin (decreasing).
Narrower bins appear earlier in the sorted list.
If two bins have the same width, they are sorted by the number of items they contain.
This sorting produces an ordered list of bins $\mathcal{B}^*$ that prioritizes narrower, denser bins.\par
Finally, to produce the transformed data $\mathcal{D}_\text{QuaRs}$, the data points within each bin are shifted to new positions based on a left-right alternation strategy.
The bins are processed sequentially in their sorted order ($\mathcal{B}^*$).
Each bin's data points are repositioned relative to a ``center'' that alternates between the left and right ends of the output space.
This alternation helps spread the bins evenly across the transformed dataset.
After processing all bins, the output dataset $\mathcal{D}_\text{QuaRs}$ contains the transformed data returned by QuaRs.
It reflects the reshuffled bin structure, with data points rearranged and repositioned to match the new bin order.\par
The decoding stage of QuaRs takes the transformed data set $\mathcal{D}_\text{QuaRs}$ and the sorted bins $\mathcal{B}^*$ as input and reverses the encoding process to restore the original data.
The first step is to reverse the bin sorting that was performed during the encoding stage.
By sorting the reshuffled bins $\mathcal{B}^*$, we obtain the bins $\mathcal{B}$ in their original order.
This step ensures that the bins are correctly aligned for the decoding process.
The next step is to reverse the positional shifts applied to the data points during encoding.
The algorithm iterates over the bins in the reshuffled order ($\mathcal{B}$) and restores the data points within each bin to their original positions.
Among other things, this involves determining the original index in the restored bin order $\mathcal{B}$, for each reshuffled bin $b_k^* \in \mathcal{B}^*$.



\section{Examples}\label{sec:results}

\begin{figure}[!ht] 
    \centering
    \includegraphics[width=\textwidth,trim={0 0cm 1cm 0},clip]{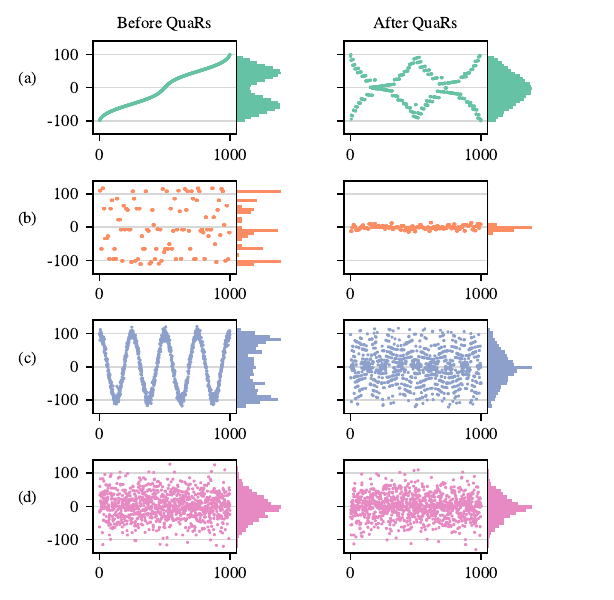} 
    \caption{Applying \gls{quars} to different numeric data sets.}
    \label{fig:quars-time-series-examples}
\end{figure}

The effect of \gls{quars} on examplary data sets can be observed in Figure \ref{fig:quars-time-series-examples}.
One can clearly observe that the effect of applying \gls{quars} is twofold: (1) the histogram of the data is transformed toward one that is unimodal and centered around zero, and (2) the average absolute value of the data is decreased.\par
The examplary data sets illustrate the impact of QuaRs (Quantile Rescaling) on different numerical distributions.
Each scenario (a-d) is represented by a pair of plots: the left column shows the data before applying QuaRs, while the right column displays the results after its application.
The results highlight the ability of QuaRs to redistribute the data into a uniform distribution that is centered around $0$.
Typically, this results in a decrease in the average absolute value of the data.

\subsection{(a) Multimodal distribution}
The initial data distribution is multimodal, with two peaks and a symmetric shape.
After applying QuaRs, the data is transformed into a unimodal distribution.

\subsection{(b) Sparse and asymmetric distribution}

The initial data distribution is sparse and asymmetric, with scattered points and a skewed shape.
QuaRs reshapes the data into a more uniform spread, with values closer to zero.

\subsection{(c) Sinusoidal pattern with noise}

The initial data distribution exhibits a sinusoidal pattern with additive white noise.
The distribution is concentrated around the peaks of the sinusoidal pattern.
This is an example of how multimodel distributions can arise in practice.
After applying QuaRs, the data is transformed into an evenly dispersed pattern.

\subsection{(d) Gaussian noise}

The initial data distribution is i.i.d. gaussian noise with a mean of zero and a standard deviation of 40.
Hence, the data is already unimodal and centered around $0$.
As a consequence, the effect of QuaRs is less pronounced.
However, we note that QuaRs has no detrimental effect on the data distribution either.

\section{Compuational complexity}\label{sec:complexity}

This section briefly describes the computational complexity of \gls{quars}, given a data set $\mathcal{D}$ of size $N$ and a number of quantiles $q$.
The quantiles can be computed in $\mathcal{O}(N \log N + q)$ operations; $\mathcal{O}(N \log N)$ for sorting the data and $\mathcal{O}(q)$ for indexing the quantile values.
The sorting of the bins requires $\mathcal{O}(q \log q)$ operations.
The transformation of the data set $\mathcal{D}$, given the sorted bins, requires $\mathcal{O}(N)$ operations.
Thus, the overall complexity of \gls{quars} is $\mathcal{O}(N \log N + q \log q)$.\par
Typically, $q \ll N$, so the complexity is dominated by the sorting of the data set, $\mathcal{O}(N \log N)$.
If the quantiles are precomputed (i.e., reused), the complexity reduces to $\mathcal{O}(N)$.



\bibliographystyle{alpha}
\bibliography{main}

\end{document}

%% file: quars_algorithm.tex
\newlength\myindent
\setlength\myindent{2em}
\newcommand\bindent{%
  \begingroup
  \setlength{\itemindent}{\myindent}
  \addtolength{\algorithmicindent}{\myindent}
}
\newcommand\eindent{\endgroup}

\SetKwBlock{Begin}{}{}

\begin{algorithm}
    \KwIn{Integer-valued data $\mathcal{D} = (d_1, \dots, d_N) \in \mathbb{Z}^N$, number of quantiles $q \in \mathbb{N}_{>0}$}
    \KwOut{Sorted bins $\mathcal{B}^*$, QuaRs-transformed data $\mathcal{D}_\text{QuaRs}$}
    \vspace{.5em}
    \hrule
    \vspace{.5em}
    \Begin(\textbf{1. Compute the q-quantiles} of the data){
        $\mathcal{Q} \leftarrow \text{quantiles}(\mathcal{D}, q)$ \\
    }
    \Begin(\textbf{2. Construct the bins}){
        $\mathcal{B} \leftarrow (b_k)_{k=1}^b = \mathcal{Q} \cup \{ q+1\ |\ q \text{ in } \mathcal{Q} \text{ more than once}\} \cup \{ \text{max}(\mathcal{D}) \}$\\
    }
    \Begin(\textbf{3. Two-level sort the bins} by){
        \textbf{(i)} width of bin (increasing)\\
        \quad Bin widths $(w_k)_{k=1}^b, \quad w_k = b_{k+1}-b_k$\\
        \textbf{(ii)} number of items in bin (decreasing)\\
        \quad Histogram $(h_k)_{k=1}^b, \quad h_k = { d \mid b_k \leq d < b_{k+1}, d \in \mathcal{D}}$\\
        Sort index $(i^*_k)_{k=1}^b \quad s.t. \quad (w_{i_k^*} < w_{i_{k+1}^*}) \lor \bigl( (w_{i_k^*} = w_{i_{k+1}^*}) \land (h_{i_k^*} \geq h_{i_{k+1}^*}) \bigr)$\\
        Sorted bins $\mathcal{B}^* = (b_{i_k^*})_{k=1}^b$\\
    }
    \Begin(\textbf{4. Reshuffle the bins}){
        $\text{left, right} \leftarrow 0$\\
        $\mathcal{D}_\text{QuaRs} \leftarrow \mathcal{D}$ (Can also initialize as empty array)\\
        \For{$k \leftarrow 1$ \KwTo $b$}{
            Alternate between left and right\\
            \If{$k$ is even}{
                $\text{pos} \leftarrow \text{right}$\\
                $\text{right} \leftarrow \text{right} + w_{i_{k}^*}$
            }
            \Else{
                $\text{left} \leftarrow \text{left} - w_{i_{k}^*}$\\
                $\text{pos} \leftarrow \text{left}$
            }
            Shift data to new position\\
            $\text{mask} \leftarrow \{j \mid b_{i_{k}^*} \leq d_{j} < b_{i_{k}^*+1},\ d_j \in \mathcal{D}\}$\\
            $\mathcal{D}_\text{QuaRs}[\text{mask}] \leftarrow \mathcal{D}[\text{mask}] + \text{pos} - b_{i_k^*} \quad$ (Pointwise arithmetic)
        }
    }    
    \caption{QuaRs (encode)}
    \label{alg:quars-encode}
\end{algorithm}

\begin{algorithm}
    \KwIn{QuaRs-transformed data $\mathcal{D}_\text{QuaRs} = (d_{\text{QuaRs}, 1}, \dots, d_{\text{QuaRs}, N})$, Sorted bins $\mathcal{B}^*$}
    \KwOut{Decoded data $\tilde{\mathcal{D}}$}
    \vspace{.5em}
    \hrule
    \vspace{.5em}

    \Begin(\textbf{1. Sort the bins to obtain the original bin order}){
        $\mathcal{B} \leftarrow (b_k)_{k=1}^b = \text{sort}(\mathcal{B}^*)$\\
    }
    \Begin(\textbf{2. Invert the reshuffling}){
        $\text{left, right} \leftarrow 0$\\
        $\tilde{\mathcal{D}} \leftarrow \mathcal{D}_\text{QuaRs} \quad$ (Can also initialize as empty array)\\ 
        Iterate over shuffled bins $\mathcal{B}^*$\\
        \For{$k \leftarrow 1$ \KwTo $b$}{
            Original index $i \leftarrow$ index of $b_k^*$ in $\mathcal{B}$\\
            Bin width $w \leftarrow b_{orig_idx+1} - b_{orig_idx}$\\
            Alternate between left and right\\
            \If{$k$ is even}{
                $\text{pos} \leftarrow \text{right}$\\
                $\text{right} \leftarrow \text{right} + \text{width}$
            }
            \Else{
                $\text{left} \leftarrow \text{left} - \text{width}$\\
                $\text{pos} \leftarrow \text{left}$
            }
            Shift data to new position\\
            $\text{mask} \leftarrow \{j \mid \text{pos} \leq d_{\text{QuaRs}, j} < \text{pos} + \text{width}\}$\\
            $\tilde{\mathcal{D}}[\text{mask}] \leftarrow \mathcal{D}_\text{QuaRs}[\text{mask}] - \text{pos} + b_{k} \quad$ (Pointwise arithmetic)
        }
    }
\caption{QuaRs (decode)}
\label{alg:quars-decode}
\end{algorithm}